\documentstyle[prl,aps]{revtex}

\def\eq{\begin{equation}}
\def\eeq{\end{equation}}
\def\eqa{\begin{eqnarray}}
\def\eeqa{\end{eqnarray}}

\def\ol#1{\overline{#1}}
\def\ssz{{\scriptscriptstyle Z}}
\def\ssw{{\scriptscriptstyle W}}
\def\mh{m_h}
\def\twi#1{\tilde{#1}}

\begin{document}
\title{A Higgs or Not a Higgs?\\
\vspace{-2cm}
\rightline{\rm McGill-00/29, IASSNS-HEP-00/72}
\vspace{2cm}}
\author{C.P. Burgess,${}^{1,2}$ J. Matias${}^3$ and M. Pospelov${}^4$\\
\vspace{3mm}
{\small
$^{1}$ Physics Department, McGill University,
3600 University Street, Montr\'eal, Qu\'ebec, Canada H3A 2T8.}\\
{\small
$^{2}$ Institute for Advanced Study, Princeton, NJ, 08540.}\\
{\small
$^{3}$ Institut f\"ur Theoretische Physik E, RWTH Aachen, 
52056 Aachen, Germany.}\\
{\small
$^{4}$ Theoretical Physics Institute,
University of Minnesota, Minneapolis MN, USA 55455.}}

\medskip
\date{September 2000}

\maketitle
\begin{abstract}
This talk summarizes a method for analyzing the properties of
any new scalar particle, which is systematic in the sense that
it minimizes {\it apriori} theoretical assumptions about the properties 
of the scalar particle, leading to very model-independent results. 
This kind of analysis lends itself to systematic survey through the
terrain of candidate theories, which we find has vast unpopulated areas. 
It is also useful for quantifying the comparison of the goodness of fit 
of competing descriptions of data, should a new scalar be found.
\end{abstract}
\bigskip 
\bigskip

\section{Motivation}

This talk\footnote{Presented by C. Burgess to ICHEP XXX, Osaka, Japan, July/August 2000.} 
is a telegraphic summary
of the much more detailed discussion of the physics of a new scalar 
presented in ref.~[1]. We encourage interested readers to 
look to this reference, which fills in the fine pencil work behind the
broad brush strokes presented here. (Lack of space also necessarily limits the
number of papers we can cite, so please see [1] for more extensive referencing.)

Much has been written about the properties of the Higgs boson, 
both in its Standard Model (SM) guise, or within one of the more popular variant 
models, such as two doublet models (THDMs), 
left-right symmetric models (LRSMs) or supersymmetric generalizations of 
these.\cite{HiggsReviewsTH,HiggsReviewsEX}
Considerable experimental effort also has gone into Higgs searches, partly guided by the
many detailed theoretical studies. The recent indications for a Higgs having a
mass of order 115 GeV has led to an extension of LEP's running time, and may yet bring news
of a final discovery. 

But if a new scalar is indeed found, how can we know if it is our friend the Higgs rather than
some other kind of scalar imposter? Ideally, this is answered by measuring all of the scalar's
couplings and comparing the results to the well-known SM predictions. Unfortunately, the
precision required to distinguish the SM Higgs from its popular close cousins is not likely to
be available soon after discovery. 

This talk addresses what we can do in the meantime. Instead of being glum
due to the cup being half-empty -- {\it i.e.} over our likely inability to distinguish scalars 
coming from well-motivated, but closely related models --  we would like to rejoice at it being half-full:
there will be numerous theories which predict scalars which are experimentally 
distinguishable from the SM very early on. 
It was the purpose of Ref. [1] to provide the first systematic roadmap to these dark and poorly
explored corners of theory space.

\section{The Framework}\label{sec:EffCpl}

Of course any analysis must come with working assumptions, our goal is to
minimize ours and to tie them closely to physical questions. We assume
that at first only a new scalar is discovered, and all other new particles are
reasonably heavy compared to it. {\it E.g.:} if the new scalar has mass 115 GeV, 
we imagine all other particles being much heavier (say $> 200$ GeV). 
This assumption permits the analysis of the scalar's properties within the effective
theory obtained by integrating out all other heavier particles.
The lowest-dimension effective couplings of such a scalar are the most important at
low energies. Up to dimension 4 the complete list of couplings is:
$$
\label{dimtwoopsa}
{ \mh^2 \over 2} \;
h^2  + { \nu \over 3!} \; h^3  + {a_\ssz \over 2} \;
Z_\mu Z^\mu \; h +   a_\ssw \; W^*_\mu W^\mu \; h  , 
$$
and
$$
\label{dimfouropsa}
\sum_{Q(f) = Q(f')}
\ol{f} \Bigl( y_{ff'} + i \gamma_5  z_{ff'} \Bigr) 
f' \; h  + \, {\lambda \over 4!} \; h^4
 + \left( {b_\ssz \over 4} \; Z_\mu Z^\mu +
{b_\ssw\over 2} \; W^*_\mu W^\mu \right) \; h^2 
$$
Some dimension-five interactions can also be important:
\eqa
\label{dimfiveopsa}
&& c_g \; G^\alpha_{\mu\nu}
G^{\mu\nu}_\alpha \; h + \tilde{c}_g \; G^\alpha_{\mu\nu}
\twi{G}^{\mu\nu}_\alpha \; h + c_\gamma \;
F_{\mu\nu} F^{\mu\nu} \; h \nonumber\\
&& \tilde{c}_\gamma \;
F_{\mu\nu} \twi{F}^{\mu\nu} \; h +  c_{\ssz\gamma}
Z_{\mu\nu} F^{\mu\nu} \; h
+\tilde c_{\ssz\gamma} Z_{\mu\nu} 
\twi{F}^{\mu\nu} \; h . \nonumber
\eeqa
Ref.~[1] gives expressions for how observables depend on these
couplings without making common theoretically-motivated
assumptions (like $y_f \propto m_f/v \ll 1$). It also collects current experimental
limits on their size. 

\section{Consequences}

\noindent{\it 1. Map of Model Space} The kinds of experimental distinctions 
likely to be possible soon after discovery can be summarized by the answers
provided to four questions. ($i$) Q1: Are trilinear $hWW$ and $hZZ$ couplings 
of order electromagnetic in size ($O(e)$ or larger)? ($ii$) Q2: Is the same
true for Yukawa couplings? ($iii$) Q3: Are electromagnetic $h\gamma\gamma$ 
couplings $O(e^2/16\pi^2)$ or larger? ($iv$) Are gluonic $hgg$ couplings
$O(g^2/16\pi^2)$ or larger? 

There are 12 possible combinations of answers to these 4 yes/no questions 
because a `yes' answer to Q1 generally implies a `yes' answer to Q3.
Table 1 enumerates the 12 options, and places the 
most popular models. Three features emerge: 
\begin{enumerate}
\item The most
popular models tend to cluster together, making them difficult to easily
distinguish from one another. 
\item Models are not {\it completely} clustered
so experiments can immediately provide {\it some} information about the
viability of {\it some} popular models. 
\item Some categories are empty in
Table 1, indicating a failure of theoretical imagination. Should experiments
point us to the empty slots, theorists will fill them, so we must bear in
mind they can exist. 
\end{enumerate}

Similarly general statements may be made concerning the finer distinctions
amongst models sharing one of the entries of the Table. For instance
loop corrections to Yukawa couplings are known to distinguish supersymmetric models from some
2HDMs. We show how these arguments rely on an underlying chiral symmetry, and so
apply more generally than to these two alternatives. Alternatively, by comparing
general expressions for $hWW$ and $hZZ$ couplings in multi-Higgs models, we find
general inequalities which these couplings must satisfy, depending on the electroweak
representation filled out by the various Higgses. 

\section*{Acknowledgments}
Support from  N.S.E.R.C. (Canada), F.C.A.R. (Qu\'ebec), DoE Grant No. DE-FG02-94ER40823 (US), 
the Ambrose Monell Foundation and a Marie 
Curie EC grant (TMR-ERBFMBICT 972147) is gratefully acknowledged.

\newpage

\begin{table*}[t]
\begin{center}
{\bf Table 1}\\[6pt]
\begin{tabular}{c|p{200pt}|cccc}  
\hline
\raisebox{0pt}[16pt][6pt]{Class} &Examples & Q1 & Q2 & Q3 & Q4 \\[3pt]
\hline 
\raisebox{0pt}[16pt][6pt]{I} & SM, 2HDM (+), LRSM (+), SUSY (+) 
 & Y & Y & Y & Y  \\[6pt] 
\hline
\raisebox{0pt}[16pt][6pt]{II} & Triplet ($\nu$,+) & Y & Y & Y & N  \\[6pt] 
\hline
\raisebox{0pt}[16pt][6pt]{III} &	& Y & N & Y & Y  \\[6pt] 
\hline
\raisebox{0pt}[16pt][6pt]{IV} & TechniPGBs, LRSM ($-$), 2HDM ($-$), 
SUSY ($-$)  & N & Y & Y & Y  \\[6pt] 
\hline
\raisebox{0pt}[16pt][6pt]{V} & Higher Representation (+) & Y & N & Y & N \\[6pt] 
\hline
\raisebox{0pt}[16pt][6pt]{VI} & Triplet ($\nu,-$) & N & Y & Y & N  \\[6pt] 
\hline
\raisebox{0pt}[16pt][6pt]{VII} &	& N & Y & N & Y  \\[6pt] 
\hline
\raisebox{0pt}[16pt][6pt]{VIII} &	& N & N & Y & Y  \\[6pt] 
\hline
\raisebox{0pt}[16pt][6pt]{IX} & Singlet w. RH $\nu$ ($\nu$), Cons. Q.No. ($\nu$) & N & Y & N & N  \\[6pt] 
\hline
\raisebox{0pt}[16pt][6pt]{X} &	& N & N & Y & N  \\[6pt] 
\hline
\raisebox{0pt}[16pt][6pt]{XI} &	& N & N & N & Y  \\[6pt] 
\hline
\raisebox{0pt}[16pt][6pt]{XII} & Higher Representation ($-$) & N & N & N & N  \\[6pt] 
\hline
\end{tabular}
\medskip\caption{ The twelve categories of models,
based on the size of their effective couplings. The positions
of some representative models are indicated, where CP conserving
scalar couplings are assumed for simplicity. ($\pm$) denotes
the CP quantum number of the observed light scalar state. A $\nu$
in brackets indicates that the large Yukawa coupling may
be restricted to neutrinos only. Categories XIII through XVI
are not listed because models having $O(e)$ couplings to
the $W$ and $Z$ generally also have $O(\alpha/2 \pi)$
effective couplings to photons. Triplet indicates a doublet-triplet
model for which the observed light scalar is dominantly from the
triplet component.\label{tab:exp}}
\end{center}
\end{table*}


\begin{thebibliography}{99}
%
\bibitem{HoNH}
{\it A Higgs or Not a Higgs? What to Do if You Discover a New Scalar
Particle}, C.P. Burgess, J. Matias and M. Pospelov, (hep-ph/9912459).
%
\bibitem{HiggsReviewsTH}
Theoretical discussions may be found in the contributions
of Klaus Desch and Jan Kalinowski to this volume.
%
\bibitem{HiggsReviewsEX}
News about experimental Higgs searches in this volume are by 
Alessandra Caner, Ian Fisk, Peter Igo-Kemenes, Shan Jin, Ari Kiiskinen, 
Wolfgang Lohmann, Kaori Maeshima and Maria Roco. 
%
\end{thebibliography}
\end{document}